\documentclass[prb,superscriptaddress,twocolumn,floatfix,amsmath,amssymb,showpacs]{revtex4}
\usepackage{graphicx}
\usepackage{dcolumn}
\usepackage{verbatim}
\usepackage{times}
\usepackage{subfigure}
\usepackage{bm}
\usepackage{color}
\usepackage{subfigure}
\usepackage[colorlinks,bookmarks=false,citecolor=blue,linkcolor=red,urlcolor=blue,backref=red,dvipdfm]{hyperref}
\usepackage{booktabs}
\usepackage{appendix}

\begin{document}

\title{Valence transition in topological Kondo insulator}

\author{Jia-Tao Zhuang}
\affiliation{College of Science, Guilin University of Technology, Guilin 541004, China}

\author{Xiao-Jun Zheng}
\affiliation{College of Science, Guilin University of Technology, Guilin 541004, China}

\author{Zhi-Yong Wang}
\affiliation{College of Science, Guilin University of Technology, Guilin 541004, China}

\author{Xing Ming}
\affiliation{College of Science, Guilin University of Technology, Guilin 541004, China}

\author{Huan Li}
\email{lihuan@glut.edu.cn}
\affiliation{College of Science, Guilin University of Technology, Guilin 541004, China}

\author{Yu Liu}
\email{liu\_yu@iapcm.ac.cn}
\affiliation{Institute of Applied Physics and Computational Mathematics, Beijing 100088, China}
\affiliation{Software Center for High Performance Numerical Simulation, China Academy of Engineering Physics, Beijing 100088, China}

\author{Hai-Feng Song}
\affiliation{Institute of Applied Physics and Computational Mathematics, Beijing 100088, China}
\affiliation{Software Center for High Performance Numerical Simulation, China Academy of Engineering Physics, Beijing 100088, China}

\date{\today}

\begin{abstract}

We investigate the valence transition in three-dimensional topological Kondo insulator through slave-boson analysis of periodic Anderson model. By including the effect of intra-atomic Coulomb
correlation $U_{fc}$ between conduction and local electrons, we find a first-order valence transition from Kondo region to mixed valence upon ascending of local level above a critical $U_{fc}$, and this valence transition usually occurs very close to or simultaneously with a topological transition.
Near the parameter region of zero-temperature valence transition, rise of temperature can generate a thermal valence transition from mixed valence to Kondo region, accompanied by a first-order topological transition.
Remarkably, above a critical $U_{fc}$ which is considerable smaller than that generating paramagnetic valence transition, the original continuous antiferromagnetic transition is shifted to first order one, at which a discontinuous valence shift takes place.
Upon increased $U_{fc}$, the paramagnetic valence transition approaches then converges with the first-order antiferromagnetic transition, leaving an significant valence shift on the magnetic boundary.
The continuous antiferromagnetic transition, first-order antiferromagnetic transition, paramagnetic valence transition and topological transitions are all summarized in a global phase diagram.
Our proposed exotic transition processes can help to understand the thermal valence variation as well as the valence shift
around the pressure-induced magnetic transition in topological Kondo insulator candidates and in other heavy-fermion systems.

\end{abstract}



\maketitle

\section{Introduction}

Since the proposal of topological intrinsic in some special Kondo insulators (KI)~\cite{Dzero10,Dzero12}, renewed attention has been attracted on these old materials, now are known as "topological Kondo insulators"(TKI) represented by SmB$_6$~\cite{Tran12,Alexandrov13,Legner14,Werner14,Roy14,Kim14,Baruselli14,
Li14,Xu14,Alexandrov15,Zheng18,Pirie18,Ying18,Broekey18,Ohtsubo19,Fuhrman19}. In these TKIs, the strong spin-orbit coupled hybridization between $d$ and $f$ electrons guarantees time-reversal symmetry (TRS) and generates a band inversion between $d$ and $f$ orbits at certain time-reversal-invariant momenta (TRIM), leading to a topologically protected state classified by $Z_2$ invariants~\cite{Dzero10,Dzero12,Dzero13,Alexandrov13,Dzero16}.
Via the spin- and angle-resolved photoemission spectroscopy
(SARPES), the metallic surface states with Dirac points have been observed in SmB$_6$, confirming its topological characteristic~\cite{Xu14,Xu16}.
A Variety of theoretical works have been carried out for TKI to reveal its rich topological phases, topological transitions, surface states, and magnetic transitions, through first-principle calculations as well as model calculations for periodic Anderson model (PAM)~\cite{Tran12,Lu13,Legner14,Werner14,Roy14,Kim14,Baruselli14,Legner15,Yu15,Alexandrov15}.

Recently, the high-pressure studies of SmB$_6$ demonstrate a magnetic transition at 6$\sim$8 GPa
~\cite{Barla05,Derr06,Derr08,Paraskevas15,Emi16,Butch16,Zhou17},
around which the mean valence $\nu$ of Sm ion (Sm$^{\nu+}$) displays a
quite rapid variation from 2.5 to saturated value 3~\cite{Zhou17}, in addition, the mean
valence also increases as temperature rises~\cite{Mizumaki09,Wu17},
similar to other TKI candidates such as YbB$_{12}$ and pressured golden
SmS~\cite{Annese06,Li14-PRB,Hagiwara17}. In the hole representation
for the $f$ shell (in the filled 4$f^6$ base) of Sm atom, the valence $\nu$ is related to
the mean occupation of $f$ holes $n_f$ by $\nu=2+n_f$~\cite{Derr06,Alexandrov13}, so the
valence shift of Sm ion indicates a variation of $f$-occupation with pressure, from mixed
valence (MV), to Kondo region (or local moment region) in which the $f$ electrons are nearly
localized to create a magnetic order. In this context, the possible valence transition or
valence crossover in TKI, as well as the relation to the magnetic transition deserve further
theoretical investigation.

For Ce- and Yb- based heavy-fermion compounds, the first-order valence transition (FOVT) was observed decades ago in pressure studies~\cite{Felner85,Mito03,Park06},
in which the valences of Ce and Yb ion increase abruptly at a critical temperature, and under enhanced pressure, FOVT can be suppressed and terminates at a critical end point (CEP) to become a valence crossover.
The valence of Ce (Yb) is manipulated by the electron (hole) occupation number $n_f$ of the 4$f$ shell~\cite{Watanabe09}, so the discontinuous valence jump indicates
a first-order transition from MV (with small $n_f$) to Kondo region (nearly localized $f$ electrons with $n_f\sim1$). Pressure can drive magnetic transitions in some heavy-fermion
compounds, and it's found that FOVT reaches the magnetic boundary (e.g. in YbInCu$_4$~\cite{Immer97,Mito03}), implying a strong interplay between FOVT and magnetic transition.
The zero-temperature FOVT in heavy-fermion systems can be understood through PAM by considering the on-site Coulomb repulsion $U_{fc}$ between local $f$
and conducting $c$ electrons~\cite{Goltsev01,Watanabe06}. Upon ascent of the energy level $\epsilon_f$ of $f$ orbit by strengthened pressure,
the influence of $U_{fc}$ causes a much rapider ascent of the renormalized $f$ level (relative to the chemical potential), pours the electrons into conduction band then consequently drives
a decrease of $f$ occupation number $n_f$, leading to the crossover behavior of the valence. Above a critical $U_{fc}$, the valence
crossover is strengthened then finally changed into FOVT from Kondo region to MV, showing a abrupt fall of $n_f$ at a critical $\epsilon_f$~\cite{Watanabe08,Watanabe09}.

Beside $\epsilon_f$, pressure applied on heavy-fermion compounds can lead to variations of other model parameters such
as the hybridization strength, which all can affect the FOVT~\cite{Watanabe09}, so the experimentally observed FOVT is a combined outcome
within model description. In YbInCu$_4$, the enhanced pressure drives a FOVT from MV to Kondo region~\cite{Mito03}, contrary to Ce-based
systems, therefore the pressure-induced FOVT cannot be simply attributed to the ascent of $\epsilon_f$~\cite{Goltsev01}. By contrast,
the FOVT from MV to Kondo region in Ce- and Yb- compounds by rising temperature is a purer effect thus can be interpreted more straightforwardly,
but such thermal FOVT is still lacking of theoretical investigation.
On the other hand, for SmB$_6$ and TKI candidate golden SmS, near the pressure-driven magnetic boundary, the valence of Sm ion shows an active increase~\cite{Zhou17,Barla04,Butch16},
similarly, at low temperatures, the pressure-induced magnetic transition in YbInCu$_4$ holds simultaneously a FOVT~\cite{Mito03}, therefore,
the relation of FOVT or valence crossover in TKI (and also in other heavy-fermion systems) to the magnetic transition should be
clarified in an unified framework. Particularly, in TKI, variation of model parameters can produce various topological phases and
distinct transition processes among them, and can driven magnetic transition as well~\cite{Li18,Li18-2}, so interest questions arise, as
how the FOVT or valence crossover appears in TKI? what is the relation between FOVT, topological transitions and magnetic transition?
can any unrevealed novel transition process takes place in TKI?

This work is devoted to answer above questions. We first study the zero-temperature valence transition in three-dimensional TKI,
modeling by spin-orbit coupled PAM with $U_{fc}$ interactions.
Similar to other typical KIs, we find the rise of $f$ energy level $\epsilon_f$ can
induce a rapid decrease of $f$ electron number $n_f$, causing a valence crossover of $f$ orbit, then above a critical $U_{fc}$,
the valence crossover is shifted to a FOVT from Kondo region to MV. We also find that the $U_{fc}$ interaction has an significant
impact on the topological boundaries of TKI. Remarkably, FOVT generally takes place very close to a topological boundary, and with slightly
greater $U_{fc}$ than the critical one, FOVT can simultaneously cause a first-order topological transition, which should be clarified
more rigorously in future studies. We also propose a thermal FOVT of TKI from MV to Kondo region by rising temperature in a narrow
parameter regime, simultaneously with a first-order topological transition. Furthermore, we find the effect of $U_{fc}$ interaction
leads to a strong variation of $f$ valence near the continuous antiferromagnetic (AF) transition in TKI, and further enhancement of
$U_{fc}$ can drive the AF transition from continuous one to first order one, meanwhile, the continuous valence variation is shifted
to a FOVT at the AF boundary. In a narrow $U_{fc}$ window, we find a gradual approaching then convergence between paramagnetic (PM) FOVT and
FOVT-associated first-order AF transition on $U_{fc}$-$\epsilon_f$ plane.
In addition, such FOVT-associated first-order AF transition can also be generated by increased temperature in some parameter
regime. The PM FOVT, two classes of AF transitions and the topological transitions are all summarized in a global phase diagram.

Our work provides new insight into the exotic transition precesses in TKI, and can be used to qualitatively understand the observed thermal valence variation as well as
the valence change around the magnetic transition in SmB$_6$, other TKIs and heavy-fermion systems.

\section{valence transition and topological transitions}
We use the spin-1/2 half-filled PAM in cubic lattice with a spin-orbit coupled $c$-$f$ hybridization to describe TKI~\cite{Dzero12,Legner14}.
This model is adopted frequently in the literature and successfully reveals the topological aspects of TKI~\cite{Werner14,Alexandrov15,Li18,Peters18}.
Besides, the on-site $c$-$f$ Coulomb interaction $U_{fc}$ is the crucial driving force of FOVT in heavy-fermion systems thus should be included~\cite{Goltsev01,Watanabe08,Watanabe09}. The model Hamiltonian of PAM reads:
\begin{align}
\mathcal{H}=&\sum_{i,j,\sigma}(-t^c_{ij}c^\dag_{i\sigma}c_{j\sigma}-t^f_{ij}f^\dag_{i\sigma}f_{j\sigma})
+\epsilon_f\sum_{i,\sigma}f^\dag_{i\sigma}f_{i\sigma}\nonumber\\
+&U\sum_in^f_{i\uparrow}n^f_{i\downarrow}
-(\frac{\textrm{i}V}{2}\sum_{i,\vec{l},\alpha,\beta}\vec{l}\cdot\vec{\sigma}_{\alpha\beta}c^\dag_{i\alpha}f_{i+\vec{l},\beta}+h.c.)\nonumber\\
+&U_{fc}\sum_in^c_in^f_i-\mu\sum_{i,\sigma}(c^\dag_{i\sigma}c_{i\sigma}+f^\dag_{i\sigma}f_{i\sigma}),
\label{PAM}\end{align}
in which half-filling of total electrons $n_t=n_c+n_f=2$ is fixed by adjusting the chemical potential $\mu$,
$\vec{\sigma}$ is three-dimensional vector formed by three Pauli matrices. We choose the electron hopping amplitude (EHA)
up to next-next-nearest-neighbor, with the two sets of EHAs shown in Tab.\ref{EHA}, both retaining an
insulating bulk gap~\cite{Li18}. The spin- and oriental- dependent hybridization between $c$ and $f$ electrons on
neighboring sites linking by coordination vector $\vec{l}$ guarantees the TRS of above PAM.
The on-site $f$-$f$ Coulomb repulsion $U$ is general large thus is set to infinite for simplicity~\cite{Alexandrov13,Alexandrov15,Li18,Peters18},
appropriate for applying the standard slave-boson mean-field technique, through which we arrive
at the effective Hamiltonian in momentum space as
\begin{align}
&H_{\mathrm{MF}}=N[\lambda(b^2-1)-U_{fc}n_cn_f]+\nonumber\\
&\sum_{\mathbf{k},\alpha,\beta}(c^\dag_{\mathbf{k}\alpha},f^\dag_{\mathbf{k}\alpha})
\left(
\begin{array}{cc}
\tilde{\epsilon}^c_\mathbf{k}\delta_{\alpha\beta}&\tilde{V}\mathbf{S}_{\mathbf{k}}\cdot \vec{\sigma}_{\alpha\beta}\\
\tilde{V}\mathbf{S}_{\mathbf{k}}\cdot \vec{\sigma}_{\alpha\beta}&\tilde{\epsilon}^f_\mathbf{k}\delta_{\alpha\beta}
\end{array}
\right)
\left(\begin{array}{cc}
c_{\mathbf{k}\beta}\\f_{\mathbf{k}\beta}
\end{array}
\right),
\label{PAM-2}\end{align}
in which $\alpha$ and $\beta$ represent spin orientations, $\lambda$ is the lagrange multiplier, $b$ is the mean expectation value of slave bosons obeying the
relation $n_f=1-b^2$. The effective hybridization is renormalized as $\tilde{V}=Vb$. $U_{fc}$ term in Eq. \ref{PAM} has been decoupled
via Hatree-Fock approximation. The renormalized $c$ and $f$ dispersions are $\tilde{\epsilon}^c_\mathbf{k}=\epsilon^c_\mathbf{k}+U_{fc}n_f-\mu$
and $\tilde{\epsilon}^f_\mathbf{k}=b^2\epsilon^f_\mathbf{k}+\epsilon_f+\lambda+U_{fc}n_c-\mu$, respectively, in which
the tight-binding dispersions $\epsilon^c_\mathbf{k}$ and $\epsilon^f_\mathbf{k}$ are determined by their corresponding EHAs~\cite{Li18}.
$\mathbf{S}_\mathbf{k}=(\sin \mathbf{k}\cdot\mathbf{a}_1,\sin \mathbf{k}\cdot\mathbf{a}_2,\sin \mathbf{k}\cdot\mathbf{a}_3)$,~\cite{Alexandrov15}
where $\mathbf{a}_1$, $\mathbf{a}_2$, $\mathbf{a}_3$ are the element vectors of cubic lattice.

\heavyrulewidth=1bp

\begin{table}
\small
\renewcommand\arraystretch{1.3}
\caption{\label{EHA}
Two sets of EHA used in this work}
\begin{tabular*}{8cm}{@{\extracolsep{\fill}}ccccccc}
\toprule
     &$t_d$ & $t^\prime_d$ & $t^{\prime\prime}_{d}$  & $t_f$ & $t^\prime_f$ & $t^{\prime\prime}_{f}$ \\
\hline
EHA(I)&1 & 0.15 & 0 & -0.2 & -0.02 & 0 \\
EHA(II)&1 & -0.375 & -0.375 & -0.2 & 0.09 & 0.09 \\
\bottomrule
\end{tabular*}

\end{table}

The quasi-particle dispersions are
derived by diagonalizing the Hamiltonian matrix in Eq.\ref{PAM-2} (in its modified form) as $E^\pm_\mathbf{k}=\frac{1}{2}[\tilde{\epsilon}^c_\mathbf{k}+\tilde{\epsilon}^f_\mathbf{k}
\pm\sqrt{(\tilde{\epsilon}^c_\mathbf{k}-\tilde{\epsilon}^f_\mathbf{k})^2+4\tilde{V}^2S^2_\mathbf{k}}]$, both doubly degenerated, then the ground state energy
is expressed by $E_g=N[\lambda(b^2-1)-U_{fc}n_cn_f]+2\sum_{\mathbf{k},\pm}\theta(-E^\pm_\mathbf{k})E^\pm_\mathbf{k}$.
The mean-field parameters $\lambda$, $b$, and $\mu$ are then solved by saddle point approximation of $E_g$ via $\partial E_g/\partial \mu=-n_t$, $\partial E_g/\partial \lambda=0$ and $\partial E_g/\partial b=0$, to obtain the following set of equations
\begin{align} &n_t=\frac{2}{N}\sum_{\mathbf{k},\pm}\theta(-E^\pm_{\mathbf{k}}),\nonumber\\
&n_f=\frac{1}{N}\sum_{\mathbf{k},\pm}\theta(-E^\pm_{\mathbf{k}})
[1\mp\frac{\tilde{\epsilon}^c_\mathbf{k}-\tilde{\epsilon}^f_\mathbf{k}}{\sqrt{(\tilde{\epsilon}^c_\mathbf{k}-\tilde{\epsilon}^f_\mathbf{k})^2+4V^2b^2S^2_\mathbf{k}}}],\nonumber\\
&\lambda=-\frac{1}{N}\sum_{\mathbf{k},\pm}\theta(-E^\pm_{\mathbf{k}})
[\epsilon^f_\mathbf{k}\pm\frac{2V^2S^2_\mathbf{k}-\epsilon^f_\mathbf{k}(\tilde{\epsilon}^c_\mathbf{k}-\tilde{\epsilon}^f_\mathbf{k})}{\sqrt{(\tilde{\epsilon}^c_\mathbf{k}-\tilde{\epsilon}^f_\mathbf{k})^2+4V^2b^2S^2_\mathbf{k}}}],
\label{sbequations}\end{align}
where $\theta(-E^\pm_{\mathbf{k}})$ is an step function. The above equations should be solved by numerical iteration.

\begin{figure}[tbp]
\hspace{-0.1cm} \includegraphics[totalheight=2.5in]{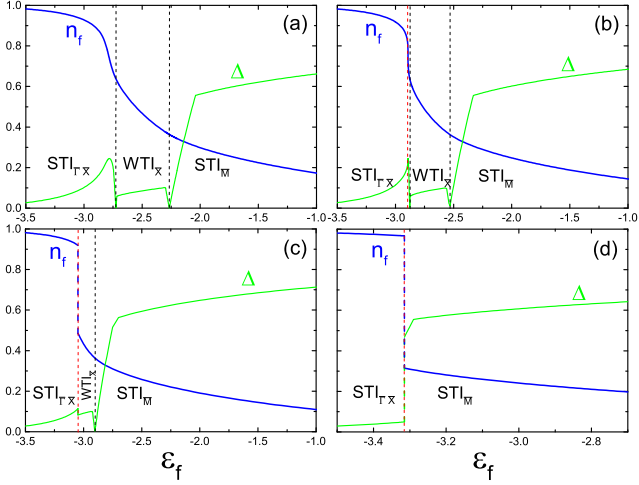}
\caption{(Color online)
Occupation number of $f$ orbit $n_f$ and bulk insulating gap $\Delta$ as functions of local level $\epsilon_f$.
From (a) to (d), $U_{fc}$=4.1, 4.31, 4.6, 5. $U_{fc}$=4.31 is the critical value, above which a FOVT occurs at a certain $\epsilon_f$,
denoted by red dashed vertical lines. When $U_{fc}$ is greater than 4.38, a first-order topological transition takes place simultaneously at FOVT
 ((c) and (d)). Black dashed vertical lines denote the conventional
topological transitions. Parameters: EHA(II) and $V=1$.
}
\label{fig1}
\end{figure}

\begin{figure}[tbp]
\hspace{-0.2cm} \includegraphics[totalheight=4.6in]{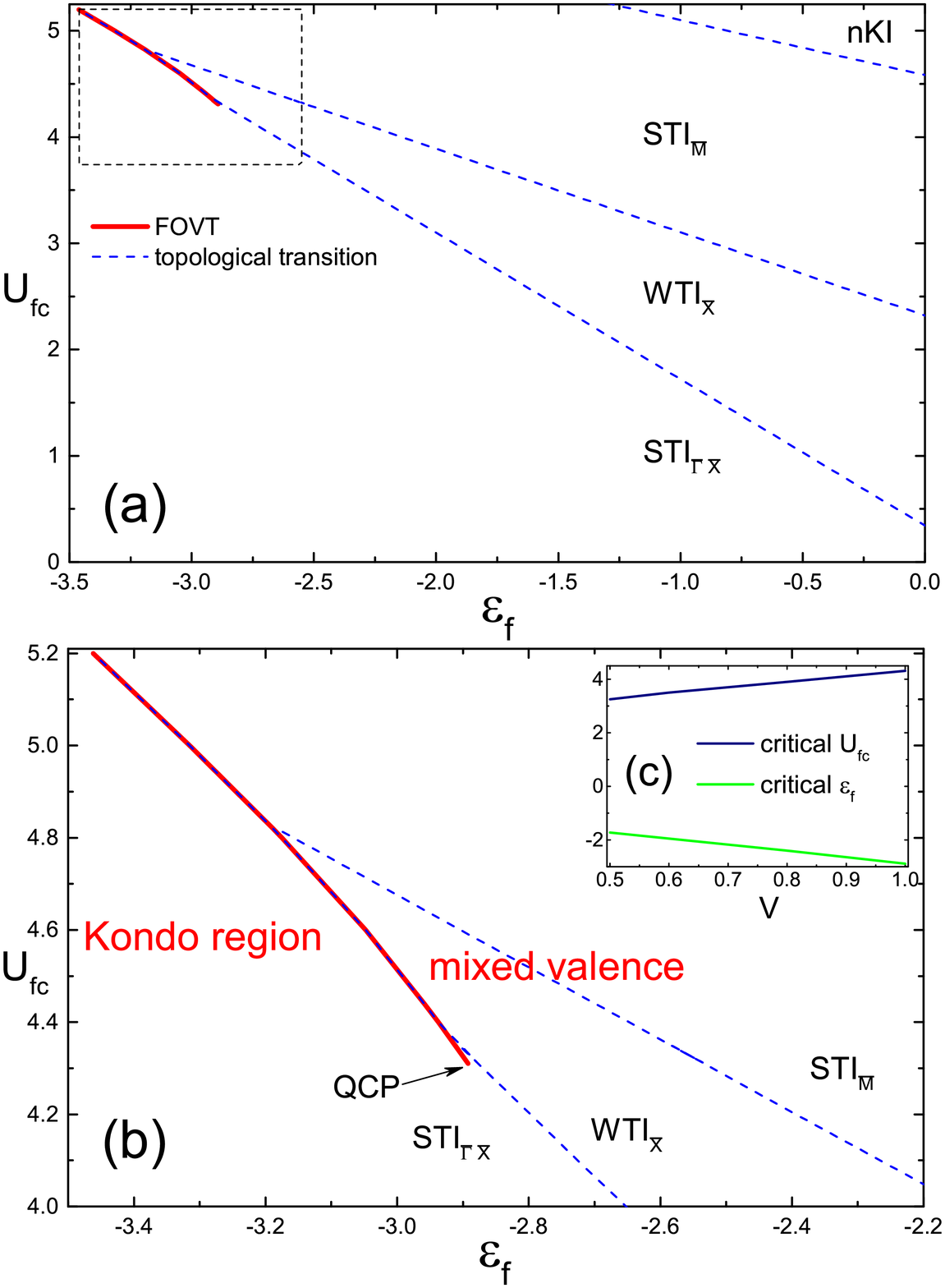}
\caption{(Color online) FOVT (red solid lines) driven by shift of local energy level $\epsilon_f$ at presence of $c$-$f$ intra-atomic Coulomb interaction $U_{fc}$.
Above a critical $U_{fc}$=4.31, a FOVT occurs between Kondo state ($n_f\sim$ 1)
and MV state (small $n_f$). Variation of $U_{fc}$ also remarkably shifts the topological boundaries
(blue dashed lines). (b) is a zoom in of the dashed area in (a). (c) illustrates the critical $U_{fc}$ and $\epsilon_f$ of QCP
as functions of hybridization $V$. Parameters: EHA(II); in (a) and (b), $V=1$.
}
\label{fig2}
\end{figure}

In SmB$_6$, the itinerant 5d band and local 4f band both locate near the Fermi level~\cite{Lu13},
causing the emergence of intra-atomic Coulomb repulsion $U_{fc}$. Similarly, in Ce and Yb systems, such
interactions are also non-ignorable and play a crucial role in their valence transitions, although the magnitude of $U_{fc}$ may differs from each other~\cite{Watanabe09}.
In this sense, the pressure- or temperature-induced valence variation of SmB$_6$ and other TKIs~\cite{Mizumaki09,Zhou17,Wu17,Annese06,Li14-PRB,Hagiwara17} should be
understood on the basis of $U_{fc}$ interaction.
On the other hand, the pressure applied on heavy-fermion systems enhances the hybridization strength $V$, and elevates the local $f$ level $\epsilon_f$ as well, therefore, to simulate the valence transition in TKI,
we first test the valence change (embodied by variation of $n_f$) as a function of $\epsilon_f$ under various magnitudes of $U_{fc}$, then study the evolution of valence transition point with $V$.

\begin{figure*}[tbp]
\hspace{-0.5cm} \includegraphics[totalheight=2.63in]{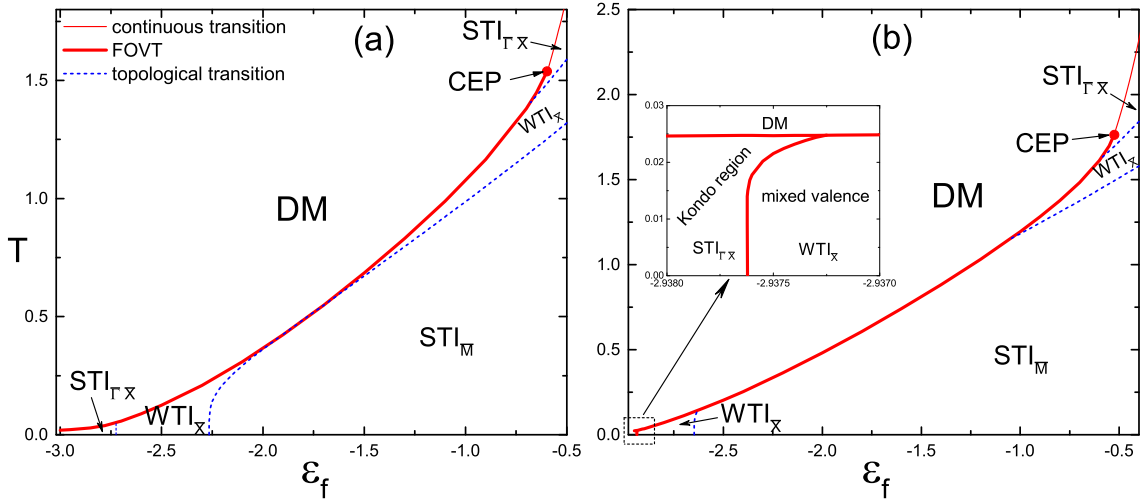}
\caption{(Color online)
Phase diagram at non-zero temperatures for (a) $U_{fc}=4.1$ and (b) $U_{fc}=4.4$. Thick red lines denote FOVT, thin red lines denote continuous decoupling of heavy-fermion state,
blue dashed lines denote topological transitions, red dots represent CEP of FOVT.
Inset of (b) shows the thermal FOVTs among MV, Kondo and DM states. Parameters: EHA(II) and $V=1$.
}
\label{fig3}
\end{figure*}
The numerical results are illustrated in Fig.\ref{fig1} with EHA(II) and $V=1$. With weak $U_{fc}$, $n_f$ decreases gradually with ascending $\epsilon_f$, showing a valence crossover behavior,
see Fig.\ref{fig1}a. While $U_{fc}$ is enhanced, $n_f$ shows a stronger decline.
At a critical value $U_{fc}=4.31$, $n_f$ drops sharply at $\epsilon_f=-2.89$, indicating the emergence of a quantum critical point (QCP), see Fig.\ref{fig1}b. While $U_{fc}$ is further increased, $n_f$ shows an abrupt jump from $n_f\sim1$ to $n_f\ll1$ at a critical $\epsilon_f$,
generating a FOVT from Kondo region in which local moments weakly hybridize with $c$ electrons, to MV region in which $c$ and $f$ electrons are strongly coupled, see Fig.\ref{fig1}c-d.
The FOVT is generated by strong $U_{fc}$ interaction which forces the $f$ electrons to pour into $c$ band when the effective $f$ level is suddenly lifted during the ascent of $\epsilon_f$.

Previous works have shown that the variation of $\epsilon_f$ or $V$ can produce successive topological transitions driven by closing and reopening of bulk insulating gap in TKI~\cite{Tran12,Legner14,Li18-2},
 and through present calculations, we find that these topological boundaries can be shifted sensitively by $U_{fc}$ towards lower $\epsilon_f$ direction, see the black dashed lines in
 Fig.\ref{fig1}a-c and blue dashed lines in Fig.\ref{fig2}. Remarkably, attributing to the rapid variation of bulk gap around
 the QCP of FOVT, the QCP usually locates very close to a certain topological transition, see Fig.\ref{fig1}b. Upon further increased $U_{fc}$ from the QCP,
 the topological boundary first approaches then finally converges with FOVT, leaving a first-order topological transition characterized by discontinuous change of insulating gap.
 The unconventional first-order topological transition is a direct outcome of FOVT in TKI, and should be verified more rigorously in future work.

 The QCP, FOVT and topological transitions are summarized on $U_{fc}$-$\epsilon_f$ plane from $U_{fc}$=$0$ in Fig.\ref{fig2}a. In Fig.\ref{fig2}b, the phase diagram around the QCP is shown in
 enlarged pattern, in which the FOVT successively converges with the topological boundaries at two points, one at $U_{fc}$=$4.38$ and $\epsilon_f$=$-2.93$, the other at $U_{fc}$=$4.82$ and $\epsilon_f$=$-3.19$,
 separating the FOVT into three segments: (1) when 4.31$<U_{fc}<$4.38, FOVT occurs near conventional STI$_{\bar{\Gamma}\bar{X}}$-WTI$_{\bar{X}}$ transition (
the subscripts denote the Dirac points on the surface Brillouin zone~\cite{Li18-2}); (2) when 4.38$<U_{fc}<$4.82, FOVT occurs
  simultaneously with first-order STI$_{\bar{\Gamma}\bar{X}}$-WTI$_{\bar{X}}$ transition; (3) when $U_{fc}>$4.82, a first-order STI$_{\bar{\Gamma}\bar{X}}$-STI$_{\bar{M}}$ transition takes place at FOVT.
Below the QCP, namely when $U_{fc}<$4.31, $U_{fc}$ is insufficient to produce FOVT, but rather a valence crossover between Kondo region and MV. Therefore, whether a heavy-fermion system
undergoes a FOVT mainly depends on the strength of $U_{fc}$, e.g., in SmB$_6$ and SmS, the pressure- and temperature- induced valence shift in PM phase seem like a crossover behavior~\cite{Butch16,Mizumaki09,Zhou17},
which may imply a weak $U_{fc}$.
It is also found that the critical $U_{fc}$ of QCP mainly depends on the hybridization strength $V$. Due to the enhanced coherence between
localized and itinerant electrons upon the increase of $V$, stronger $U_{fc}$ is required to develop a QCP of valence transition, and the critical $\epsilon_f$ is pushed to deeper direction,
see Fig.\ref{fig2}c.

\section{thermal valence transitions}
The ground-state FOVT discussed in above section may be experimentally realizable through external pressure, e.g. in Ce and Yb systems~\cite{Felner85,Mito03,Park06}, however,
the alteration of external factors will lead to
a series of changes for model parameters in PAM which are hard to track, hence it is difficult to compare above theoretical
results with experiments. Therefore, in this section, we turn to the possible valence transition caused by temperature variation in TKI,
which can be more clearly interpreted theoretically. Such "thermal valence transition"
has been detected in Ce and Yb systems decades ago, in which the increase of temperature causes a FOVT from MV to Kondo state~\cite{Felner85,Park06,Watanabe09},
but the mechanism and transition process of thermal valence transition are still lacking of theoretical verification.
In SmB$_6$ and TKI candidates such as golden SmS, YbB$_6$ and YbB$_{12}$,
the valences of Sm and Yb both increase with temperature at ambient pressure, and in some systems, the valence even reaches the saturated value 3~\cite{Annese06,Li14-PRB,Mizumaki09,Wu17,Butch16,Hagiwara17}, but
it seems no discontinuous valence shift was observed in PM phase. We expect that the thermal valence behavior in TKI can be dramatically alternated
under different magnitudes of pressure, since the pressure applied can affect the interaction strengths in heavy-fermion systems and
lead to significant change of valence in the ground state~\cite{Annese06,Butch16,Zhou17,Hagiwara17}.

At finite temperatures, the quasi-particle spectrums share the same formulas as those of the ground state, except that the mean-field
parameters are now temperature-dependent, and are determined by the saddle-point equations derived from the free energy
 $F=N[\lambda(b^2-1)-U_{fc}n_cn_f]-2T\sum_{\mathbf{k},\pm}\ln(1+e^{-E^\pm_\mathbf{k}/T})$ by $\partial F/\partial \mu=-n_t$, $\partial F/\partial \lambda=0$ and $\partial F/\partial b=0$, giving rise to
the equation set the same as the ground state one (Eq.\ref{sbequations}), except that now the step function $\theta(-E^\pm_{\mathbf{k}})$ should be replaced by
Fermi distribution function
$f^\pm_{\mathbf{k}}=1/(1+e^{E^\pm_\mathbf{k}/T})$ at non-zero temperatures.

The $T$-$\epsilon_f$ phase diagrams are given in Fig.\ref{fig3}a and \ref{fig3}b, with two magnitudes of $U_{fc}=4.1$ and $4.4$, which are slightly smaller and greater than the
QCP value $U_{fc}=4.31$, respectively. As temperature increases, $b$ is reduced gradually, but only when $\epsilon_f$ exceeds a certain value (-0.6 and -0.525 for $U_{fc}=4.1$, $4.4$, respectively), $b$ (consequently $\tilde{V}$) gradually vanishes at a critical temperature, results in
an second-order insulator-metal transition from heavy-fermion MV insulating state to a decoupled metallic state (DM) in which $n_f=1$ with completely localized $f$ electrons.
Surprisingly, when $\epsilon_f$ is further reduced, the MV-DM transition becomes first-order, indicating a thermal FOVT, at which $b$ and $n_f$ undergo an abrupt jump to 0 and 1, respectively,
Such discontinuous thermal decoupling
of heavy-fermion state and MV-DM thermal FOVT accompanied by an insulator-metal transition deserves further verification.

\begin{figure}[tbp]
\hspace{-0.1cm} \includegraphics[totalheight=1.93in]{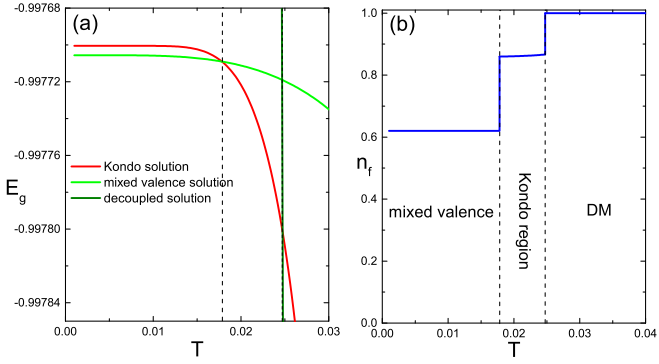}
\caption{(Color online)
Thermal MV-Kondo and Kondo-DM FOVTs in the inset of Fig.\ref{fig3}(b), with $\epsilon_f=-2.9376$. (a) Energy comparison of three solutions.
(b) $f$-electron number $n_f$ with increasing temperature, in which the dashed vertical lines denote two thermal FOVTs.
}
\label{fig4}
\end{figure}

Remarkably, when $U_{fc}$ is greater than that of the QCP, we find that with $\epsilon_f$ slightly above that triggers the zero-temperature FOVT,
there are two successive FOVTs with increasing temperature, first MV-to-Kondo FOVT, then Kondo-to-DM FOVT, see Fig.\ref{fig4}, with the transition boundaries on $T$-$\epsilon_f$ plane shown in the inset of Fig.\ref{fig3}b. Similar to zero-temperature case, the
FOVT between MV and Kondo states also holds a first-order WTI$_{\bar{X}}$-STI$_{\bar{\Gamma}\bar{X}}$ topological transition. As $\epsilon_f$ rises, the Kondo region gradually shrinks, then
two thermal FOVTs converge, leaving only a MV-DM transition. Moreover, with $\epsilon_f$ lower than that of zero-temperature FOVT, only Kondo-DM FOVT occurs with increased temperature.
For $U_{fc}$ less than the QCP value, no thermal MV-Kondo FOVT appears, since it is insufficient to support zero-temperature FOVT between MV and Kondo states.

In Ce systems, pressure can suppress the thermal FOVT, forces it to terminate at a CEP, leading to a thermal valence crossover behavior~\cite{Gschneidner78}, this feature corresponds
 to a shift of thermal behavior from first-order to continuous one. In this sense, since pressure elevates $\epsilon_f$, the $\epsilon_f$-driven order change of thermal transitions may explain the appearance of CEP in Ce systems, which is marked
 by the red dots in Fig.\ref{fig3}. In contrast to Ce systems, for SmB$_6$ and TKI candidates SmS, YbB$_6$ and YbB$_{12}$,
 the Sm and Yb valences in PM phases seem to increase continuously with temperature~\cite{Annese06,Li14-PRB,Mizumaki09,Wu17,Butch16,Hagiwara17}, showing no clear signal of thermal FOVT. The absence of thermal FOVT in these TKIs may
 be ascribed to their weak $U_{fc}$ interactions or their high $f$ level $\epsilon_f$ which exceeds the CEP in Fig.\ref{fig3}, leading to a continuous valence increase with temperature.
 Previous works suggested that magnetic field can shift the CEP of FOVT in Ce and Yb systems towards lower $U_{fc}$ and higher $\epsilon_f$~\cite{Watanabe09,Watanabe09}, similarly, we expect magnetic field
 can also shift the CEP of TKI to the right side on $T$-$\epsilon_f$ plane in Fig.\ref{fig3}, and consequently recover a thermal FOVT in TKI under ambient pressure, which requires experimental verification.

\section{valence variation around the magnetic transition}

The high-pressure experiments of SmB$_6$ display a first-order transition to magnetically ordered state at
6$\sim$8 Gpa~\cite{Barla05,Derr06}, around which
the Sm valence shows a rapid increase~\cite{Zhou17}, afterwards, the Sm valence in the magnetic phase increases
smoothly with pressure then approaches pure trivalent above 10 Gpa.
For TKI candidate golden SmS~\cite{Li14-PRB}, the pressure-induced magnetic transition at about 2 Gpa also
exhibits first-order character~\cite{Barla04}, but around this transition, the valence increase of Sm in SmS is much sharper than in SmB$_6$~\cite{Butch16}.
In YbInCu$_4$, with enhanced pressure, FOVT reaches the magnetic boundary at low temperatures~\cite{Immer97,Mito03}, implying that the pressure-induced
magnetic transition can hold a FOVT simultaneously. The active valence change at and around the magnetic transition indicates an intimate relation between these two transitions in TKI and in other
heavy-fermion systems.

In order to study the valence variation around the magnetic transition in TKI, which is mostly likely an antiferromagnetic (AF) transition~\cite{Peters18,Chang18},
we adopt Kotliar-Ruckenstein (K-R) slave-boson mean-filed method~\cite{Kotliar86,Yang93,Sun93,Sun95}.
This technique produces very close results of TKI to those of the conventional
slave-boson approach, and has the advantage to include the magnetic order conveniently~\cite{Li18,Li18-2}. The main feature of K-R representation for AF phase is that
each $f$- creation and annihilation operator in $c$-$f$ hybridization and $f$-$f$ hopping is multiplied by a factor $Z_+$ or $Z_-$ with $Z_\pm=\sqrt{\frac{2(1-n_f)}{2-n_f\mp m_f}}$,
depending on the spin and sublattice of the $f$ operator, in which $m_f$ is the staggered magnetization of $f$ electrons.
Secondly, $\epsilon_f$ is shifted by a parameter $\eta$. $Z_\pm$ and $\eta$ are analogue to $b$ and $\lambda$ in
conventional slave-boson method, respectively. The K-R mean-field treatment leads to the following effective Hamiltonian
\begin{equation}
\mathcal{H}=N(hm_f-\eta n_f-U_{fc}n_cn_f)+\sum_{\mathbf{k}}\Psi _{
\mathbf{k} }^{\dag }\mathbf{H}_{\mathbf{k} }\Psi _{\mathbf{k}
 },
\label{AF}\end{equation}
in which the summation of $\mathbf{k}$ is restricted to the magnetic Brillouin zone (MBZ), $h$ is a staggered parameter associated with $m_f$,
$\Psi^\dag_{\mathbf{k}}$=$(d^\dag_{\mathbf{k}A\uparrow}$,$d^\dag_{\mathbf{k}A\downarrow}$,
$d^\dag_{\mathbf{k}B\uparrow}$,$d^\dag_{\mathbf{k}B\downarrow}$,
$f^\dag_{\mathbf{k}A\uparrow}$,$f^\dag_{\mathbf{k}A\downarrow}$,
$f^\dag_{\mathbf{k}B\uparrow}$,$f^\dag_{\mathbf{k}B\downarrow})$ is
eight-component creation operator for $c$ and $f$ electrons in sublattice and spin spaces, the Hamiltonian matrix is given by
 \begin{align}
 \mathbf{H}_{\mathbf{k}}=\left(
\begin{array}{cc}
\mathbf{H}^d_{\mathbf{k}} & \mathbf{V}_{\mathbf{k}} \\
\mathbf{V}^+_{\mathbf{k}} & \mathbf{H}^f_{\mathbf{k}}
\end{array}
\right) \label{Hk},
 \end{align}
in which \begin{align}
 \mathbf{V}_{\mathbf{k}}=V\left(
\begin{array}{cccc}
0&0&Z_-s_z&Z_+\Gamma_{\mathbf{k}}\\
0&0&Z_-\Gamma^*_{\mathbf{k}}&-Z_+s_z\\
Z_+s_z&Z_-\Gamma_{\mathbf{k}}&0&0\\
Z_+\Gamma^*_{\mathbf{k}}&-Z_-s_z&0&0
\end{array}
\right)
 \end{align}
hybridizes $c$ and $f$ electrons on neighboring sites.
\begin{align}
 \mathbf{H}^d_{\mathbf{k}}=\left(
\begin{array}{cc}
(t^\prime_c\gamma_{\mathbf{k}}+U_{fc}n_f-\mu) &  u^c_{\mathbf{k}}\\
  u^c_{\mathbf{k}} & (t^\prime_c\gamma_{\mathbf{k}}+U_{fc}n_f-\mu)
\end{array}
\right)\otimes I_2,
 \end{align}
\begin{align}
& \mathbf{H}^f_{\mathbf{k}}=
 &\left(
\begin{array}{cccc}
e^+_{\mathbf{k}}&0&Z_+Z_-u^f_{\mathbf{k}}&0\\
0&e^-_{\mathbf{k}}&0&Z_+Z_-u^f_{\mathbf{k}}\\
Z_+Z_-u^f_{\mathbf{k}}&0&e^-_{\mathbf{k}}&0\\
0&Z_+Z_-u^f_{\mathbf{k}}&0&e^+_{\mathbf{k}}
\end{array}
\right),
\end{align}
are the sublattice versions of tight-binding Hamiltonian for $c$ and $f$, respectively, in which the $f$ hopping amplitudes are renormalized by
$Z_\pm$. $I_2$ is a two-order unit matrix, $\Gamma_{\mathbf{k}}=s_x-\textrm{i}s_y$,
$\gamma_{\mathbf{k}}=-4(c_xc_y+c_xc_z+c_yc_z)$,
$e^\pm_{\mathbf{k}}=\epsilon_f+\eta+U_{fc}n_c-\mu\mp h+t^\prime_fZ^2_\pm\gamma_{\mathbf{k}}$,
$u^{c(f)}_{\mathbf{k}}=t_{c(f)}\lambda_{\mathbf{k}}+t^{\prime\prime}_{c(f)}g_{\mathbf{k}}$ with
$\lambda _{\mathbf{k}}=-2(c_x+c_y+c_z)$ and $g_{\mathbf{k}}=-8c_xc_yc_z$, where we denote $c_i=\cos k_i$ and $s_i=\sin k_i$ ($i=x,y,z$) for simplicity.

The free energy can be expressed as the expectation value of the effective Hamiltonian by \begin{align}F=&N(hm_f-\eta n_f-U_{fc}n_cn_f+\mu n_t)\nonumber\\&+\sum_{\mathbf{k},n,m}
(\mathbf{H}_{\mathbf{k}})_{nm}\langle (\Psi^\dag_{\mathbf{k}})_n(\Psi_{\mathbf{k}})_m\rangle,\label{FAF}\end{align}
where subscripts $n$ and $m$ are row or column numbers. Use the unitary transformation matrix $\mathrm{U}_{\mathbf{k}}$ extracted from numerical diagonalization of $\mathbf{H}_{\mathbf{k}}$, we have
\begin{align}\langle (\Psi^\dag_{\mathbf{k}})_n(\Psi_{\mathbf{k}})_m\rangle
=\sum^8_{i=1}(\mathrm{U}_{\mathbf{k}})^*_{ni}(\mathrm{U}_{\mathbf{k}})_{mi}f^{(i)}_{\mathbf{k}},
\end{align}
where $f^{(i)}_{\mathbf{k}}=1/(1+e^{E^{(i)}_{\mathbf{k}}/T})$ is the Fermi distribution of quasi-particles, and equals the step function $\theta(-E^{(i)}_{\mathbf{k}})$ at zero temperature.
The set of equations determining the parameters $n_f$, $m_f$, $h$, $\eta$, $\mu$ can be obtained through the zero point of the derivation of free energy (Eq.\ref{FAF}) with respect to them,
by using the formulas of matrix elements $(\mathbf{H}_{\mathbf{k}})_{nm}$ in Eq.\ref{Hk}~\cite{Li18}.

In order to give a global phase diagram including both PM and AF phases through K-R solution,
we also solve the K-R saddle-point equations for $n_f$, $\eta$ and $\mu$ in PM phase, which are similar to those in section II and III:
\begin{align} &n_t=\frac{2}{N}\sum_{\mathbf{k},\pm}f^\pm_{\mathbf{k}},\nonumber\\
&n_f=\frac{1}{N}\sum_{\mathbf{k},\pm}f^\pm_{\mathbf{k}}
[1\mp\frac{\tilde{\epsilon}^c_\mathbf{k}-\tilde{\epsilon}^f_\mathbf{k}}{\sqrt{(\tilde{\epsilon}^c_\mathbf{k}-\tilde{\epsilon}^f_\mathbf{k})^2+4V^2Z^2S^2_\mathbf{k}}}],\nonumber\\
&\eta=\frac{2Z}{N}\frac{\partial Z}{\partial n_f}\sum_{\mathbf{k},\pm}f^\pm_{\mathbf{k}}
[\epsilon^f_\mathbf{k}\pm\frac{2V^2S^2_\mathbf{k}-\epsilon^f_\mathbf{k}(\tilde{\epsilon}^c_\mathbf{k}-\tilde{\epsilon}^f_\mathbf{k})}{\sqrt{(\tilde{\epsilon}^c_\mathbf{k}-\tilde{\epsilon}^f_\mathbf{k})^2+4V^2Z^2S^2_\mathbf{k}}}],
\label{PMKR}\end{align}
in which $Z=\sqrt{\frac{2(1-n_f)}{2-n_f}}$, $\tilde{\epsilon}^f_\mathbf{k}=Z^2\epsilon^f_\mathbf{k}+\epsilon_f+\eta+U_{fc}n_c-\mu$,
$f^\pm_{\mathbf{k}}=1/(1+e^{E^\pm_\mathbf{k}/T})$ for finite temperatures and equals the step function $\theta(-E^{\pm}_{\mathbf{k}})$ at zero temperature, where $E^\pm_\mathbf{k}=\frac{1}{2}[\tilde{\epsilon}^c_\mathbf{k}+\tilde{\epsilon}^f_\mathbf{k}
\pm\sqrt{(\tilde{\epsilon}^c_\mathbf{k}-\tilde{\epsilon}^f_\mathbf{k})^2+4V^2Z^2S^2_\mathbf{k}}]$

\begin{figure}[tbp]
\hspace{-0.2cm} \includegraphics[totalheight=3.4in]{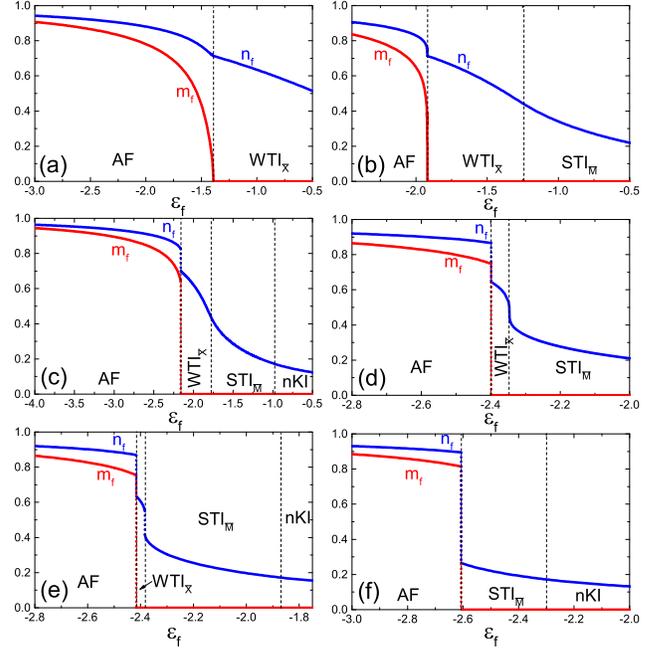}
\caption{(Color online)
Staggered magnetization $m_f$ and $f$-electron occupation $n_f$ vs local level $\epsilon_f$.
Parameters: EHA(I), $V=1$, and from (a) to (f), $U_{fc}$=2, 2.92, 3.4, 3.906, 3.94, 4.2. (a) and (b) show continuous
AF transition without valence shift. For (c) and (d), the AF transition is shifted to first-order and is accompanied by a valence jump.
For (e), the FOVT-associated AF transition and paramagnetic FOVT take place at different $\epsilon_f$, while in (f),
two transitions converge, leaving a FOVT-associated AF transition with significant valence jump.
}
\label{fig5}
\end{figure}

\begin{figure*}[tbp]
\hspace{-0.6cm} \includegraphics[totalheight=3in]{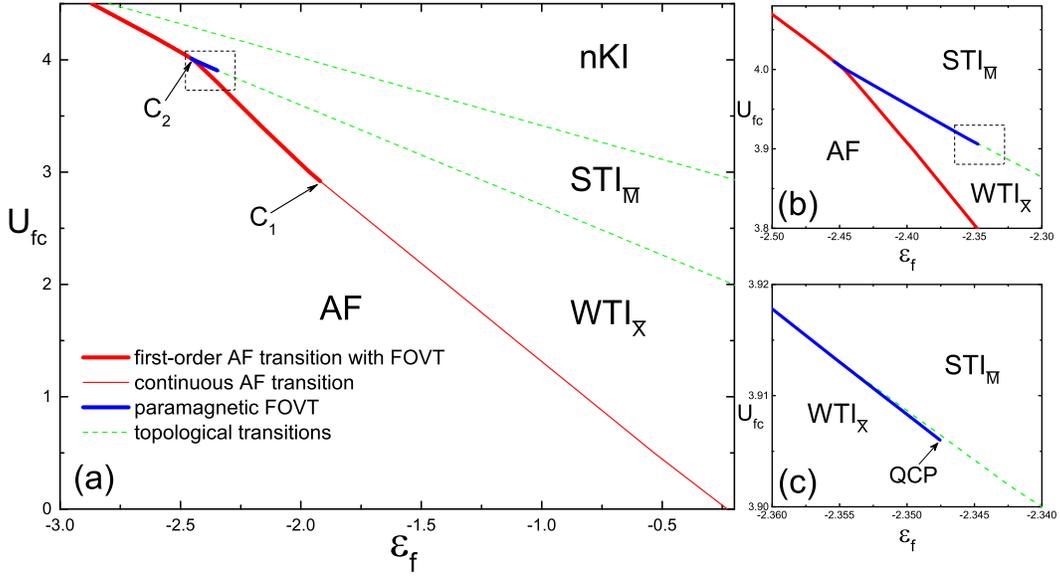}
\caption{(Color online)
$U_{fc}$-$\epsilon_f$ magnetic phase diagram, showing the continuous AF transition (thin red line),
first-order AF transition (thick red line) associating with a FOVT, and FOVT
in PM phase (thick blue line). Topological transitions are denoted by dashed green lines.
(b) and (c) are enlarged view of the dashed boxes in (a) and (b), respectively. Parameters: EHA(I), $V$=1.
}
\label{fig6}
\end{figure*}

We first study the zero-temperature phase diagram on $U_{fc}$-$\epsilon_f$ plane. For convenience, we choose EHA(I) in Tab.\ref{EHA}, set various values of $U_{fc}$ from zero, then calculate the corresponding evolution of $m_f$ and $n_f$ vs $\epsilon_f$ to locate
the AF transitions. Above the AF boundaries, the evolution of PM phases is computed via Eq.\ref{PMKR} to determine the FOVT and topological boundaries.
The results are shown with increasing $U_{fc}$ from Fig.\ref{fig5}a to \ref{fig5}f. For small $U_{fc}$, $m_f$ variation indicates an second-order magnetic transition, at which $n_f$ changes continuously with a kink;
when $U_{fc}$ exceeds 2.92, we find a first-order magnetic transition at which $n_f$ shows an abrupt jump, indicating a FOVT-associated first-order AF transition;
 in the PM phase with $\epsilon_f$ above the first-order AF boundary, $n_f$ decreases rapidly with $\epsilon_f$, and when $U_{fc}>$3.906 gives rise to a FOVT at a critical $\epsilon_f$ above the first-order AF transition; with further increased $U_{fc}$,
 the two transitions approach each other and then converge, leaving a large $n_f$ jump at the first-order AF transition.

The resulting ground-state phase diagram on $U_{fc}$-$\epsilon_f$ plane is summarized in Fig.\ref{fig6}, including the topological transitions in PM phase.
In Fig.\ref{fig6}a, it can be clearly seen that a critical point C$_1$ at ($U_{fc}=2.92,\epsilon_f=-1.92$) separates the AF boundaries into continuous part and first-order part,
which is below and above C$_1$, respectively.
The FOVT in PM phase is denoted by blue solid lines in Fig.\ref{fig6}a, and the enlarged view near the QCP at ($U_{fc}=3.906,\epsilon_f=-2.348$) is demonstrated in Fig.\ref{fig6}c,
in which the relative position between FOVT and topological boundaries behaves similarly to Fig.\ref{fig2}b.
Here we should point out that in K-R solution, the critical $U_{fc}$ and $\epsilon_f$ for the QCP of FOVT in PM phase is slightly greater and lower than the conventional
slave-boson solution, respectively, meanwhile the FOVT and topological transition processes in PM phase
 remain essential the same. The interplay between first-order AF transition and PM FOVT is clearly illustrated
in Fig.\ref{fig6}b: with increasing $U_{fc}$ from QCP, the two boundaries approach each other and finally converge at C$_2$ ($U_{fc}=4.01,\epsilon_f=-2.45$), leading to an single first-order AF boundary accompanied by a FOVT with significant valence jump.
In addition, the coexistence of FOVT-associated AF transition and PM FOVT on $\epsilon_f$ axis only occurs in a narrow $U_{fc}$ regime from 3.906 to 4.01.
Therefore, the FOVT-associated first-order AF transitions are also classified into two classes: one with weak valence shift below C$_2$, and the other with strong valence jump above C$_2$.
Consequently, the order of AF transition, as well as the "strength" of first-order AF transition both depend on the magnitude of $U_{fc}$, in this sense,
the weak first-order magnetic transitions in pressured SmB$_6$ and SmS may be ascribed to their relatively small $d$-$f$ intra-atomic Coulomb repulsion as analysed in above sections~\cite{Barla05,Derr06,Barla04},
while the remarkable valence change at the magnetic transition in YbInCu$_4$ may be attributed to its stronger $U_{fc}$~\cite{Immer97,Mito03}.
It is worth noting that the critical $U_{fc}=2.92$ of C$_1$ which triggers a first-order AF transition is much smaller than that drives a FOVT in PM phase ($U_{fc}=3.906$ of QCP),
explaining why Sm valence varies continuously with temperature at ambient pressure in SmB$_6$ and SmS, meanwhile first-order AF transitions appear under high pressure.

\begin{figure}[tbp]
\hspace{-0.1cm} \includegraphics[totalheight=2.4in]{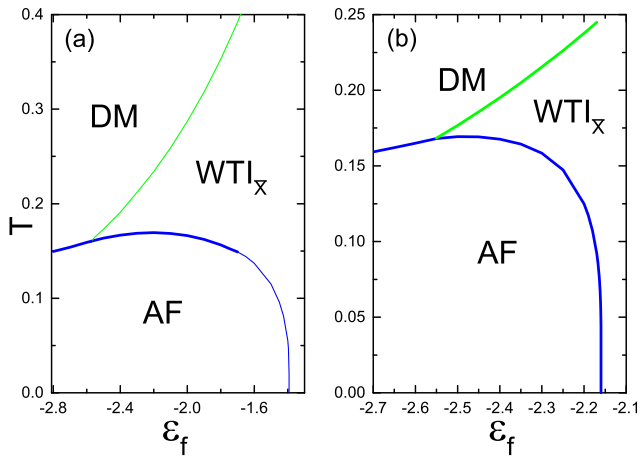}
\caption{(Color online)
AF transitions (blue lines) and transitions between heavy-fermion state and DM state (green lines).
Thin lines denote continuous transitions, while thick lines denote first-order transitions with valence shift.
Parameter: EHA(I), $V=1$.  $U_{fc}=2$ and $3.4$ in (a) and (b), respectively.
}
\label{fig7}
\end{figure}

Then we study the thermal phase transitions on $T$-$\epsilon_f$ plane, the results are displayed in Fig.\ref{fig7}. For $U_{fc}$ smaller than the critical point C$_1$, the zero-temperature
AF transition with $\epsilon_f$ is continuous, while $\epsilon_f$ descends, the thermal AF transition is shifted to first-order with valence jump (see Fig.\ref{fig7}a), implying that the critical point C$_1$ separating these two types of AF transitions is temperature dependent and
is pushed by increasing temperature towards lower $\epsilon_f$ and weaker $U_{fc}$. By comparison, for $U_{fc}$ stronger than zero-temperature C$_1$,
thermal AF transition remains first-order, because $U_{fc}$ keeps greater than C$_1$ when temperature rises. For temperature above the AF transitions,
$Z$ factor reduces with temperature, leading to the increase of $n_f$ then finally a decoupling of heavy-fermion state to DM state with $n_f=1$ at a critical temperature,
giving rise to HF-DM transition. By contrast to AF transitions, the order of this HF-DM transition is only weakly temperature-dependent and mainly depends on
$U_{fc}$: when $U_{fc}$ is less or larger than zero-temperature C$_1$ in Fig.\ref{fig6}a, HF-DM transition is continuous or first-order, respectively, and the latter coexists a FOVT. As $\epsilon_f$ descends, the AF boundary and
HF-DM boundary gradually converge, leaving a first-order AF-DM transition with abrupt valence shift.
The thermal first-order AF transition coexisting a FOVT deserves more rigorous theoretical and experimental examinations.

\section{conclusion and discussion}

To summarize, we have proposed a series of exotic valence transition processes in TKI, driven by the effect of
on-site Coulomb repulsion $U_{fc}$ between conduction and local electrons in the PAM. For the ground state, we located a QCP on the $U_{fc}$-$\epsilon_f$ plane,
above which a FOVT boundary separates the Kondo state with MV state, and holds an abrupt valence jump upon ascending $\epsilon_f$.
The QCP is very close to a topological transition, and the FOVT boundary gradually approaches then converges with the topological boundary, leading to a discontinuous topological transition.
Near the parameter regime of the zero-temperature FOVT, increase in temperature can lead to a series of thermal FOVTs among the Kondo state, MV state and DM state.
In addition, we found a CEP on $T$-$\epsilon_f$ plane which terminates the thermal FOVT from heavy-fermion state to DM state, and shift the thermal decoupling of heavy-fermion state into continuous one under higher $\epsilon_f$ above the CEP.
Further descent of $\epsilon_f$ can generate an AF order in TKI, and above a critical $U_{fc}$ which is considerably smaller than that of the QCP which triggers a PM FOVT,
the $\epsilon_f$-driven continuous AF transition is shifted to first-order, accompanied by a FOVT. Further increasing of $U_{fc}$ can push the first-order AF transition to approach then to join the PM FOVT boundary, giving rise to a
first-order AF transition with large valence shift. The thermal AF transitions can also be classified into continuous one or first-order one, depending mainly on the magnitude of $U_{fc}$.
The distinct valence-variation processes with temperature and pressure observed in TKI candidates SmB$_6$, golden SmS and other heavy-fermion compounds, as well as their valence behaviors near the
magnetic transitions can be qualitatively understood by our results in terms of their different magnitudes of $U_{fc}$. This work has used an simplified PAM with single conduction and $f$ band to
describe TKI, actually in real cubic TKI systems such as SmB$_6$, dominant hybridization channel involves spin-degenerated 5$d_{x^2-y^2}$, 5$d_{3z^2-r^2}$ orbits, and $\Gamma^{(1)}_8$, $\Gamma^{(2)}_8$ 4$f$ quartet, resulting
in much complicated tight-binding and hybridization terms~\cite{Takimoto11,Alexandrov13,Roy14}. Although the multiple parameters can all influence the detailed valence-variation process in TKI, we
believe that the intra-atomic Coulomb interaction $U_{fc}$ between $d$ and $f$ orbits still provides a crucial factor controlling the valence transition, and the magnitudes of $U_{fc}$ in TKI candidates may be examined through first-principle simulations~\cite{Lu13}.

\acknowledgments

H. Li is supported by NSFC (No. 11764010) and Guangxi Natural Science Foundation (No. 2017GXNSFAA198169).
Y. Liu and H. F. Song thank the Science Challenge Project
(Grant No. TZ2018002). Y. Liu is also supported by Open
Research Fund Program of the State Key Laboratory of Low-Dimensional Quantum Physics (No. KF201702) and the SPC-Lab
Research Fund (No. XKFZ201605). Z. Y. Wang is supported by NSFC (No. 11564008). X. J. Zheng is supported by NSFC (No.
11704084) and Guangxi NSF (No. 2017GXNSFBA198115). X. Ming is supported by NSFC (No.
11864008) and Guangxi NSF (No. 2018GXNSFAA138185).

\end{document}